\documentclass[aps,prb,twocolumn,showpacs,superscriptaddress,groupedaddress,footinbib]{revtex4}
\usepackage[dvips]{graphicx}
\usepackage{epstopdf}
\usepackage{color}
\usepackage{ucs}
\usepackage[utf8x]{inputenc}

\begin{document}

\title{The Icosahedral (H$_2)_{13}$ Supermolecule} \date{\today}
\author{Graeme~J.~Ackland} 
\affiliation{CSEC,  School of Physics and Astronomy, The University of Edinburgh,   Edinburgh EH9 3FD, UK}
\author{Jack Binns} 
\affiliation{Center for High Pressure Science Technology Advanced
Research (HPSTAR), Shanghai, China.} 
\author{Ross Howie} 
\affiliation{Center for High Pressure Science Technology Advanced
Research (HPSTAR), Shanghai, China.} \pacs{}
\author{Miguel Martinez-Canales} 
\affiliation{CSEC,  School of Physics and Astronomy, The University of Edinburgh,   Edinburgh EH9 3FD, UK}

\begin{abstract}
We investigate a range of possible materials containing the
supermolecular form of hydrogen comprising 13 H$_2$ molecules arranged
in an icosahedral arrangement.  This supermolecule consists of freely
rotating 12 H$_2$ molecules in an icosahedral arrangement, enclosing
another freely rotating H$_2$ molecule.  To date, this supermolecule
has only been observed in a compound with Iodane (HI).  The extremely
high hydrogen content suggests possible application in hydrogen
storage so we examine the possibility of supermolecule formation at
ambient pressures.  We show that ab initio molecular dynamics
calculations give a good description of the known properties of the
HI(H$_2)_{13}$ material, and make predictions of the existence of
related compounds Xe(H$_2)_{13}$, HBr(H$_2)_{13}$ and HCl(H$_2)_{13}$,
including a symmetry-breaking phase transition at low temperature. The
icosahedral (H$_2$)$_{13}$ supermolecule, remains stable in all these
compounds. This suggests (H$_2$)$_{13}$ is a widespread feature in
hydrogen compounds, and that appropriately–sized cavities could hold hydrogen
supermolecules even at low pressure.  The structure of the
supermolecule network is shown to be independent of the compound at
equivalent density.

%    Using \emph{ab initio} molecular dynamics, we have explored candidate materials that could host supermolecular (H$_2$)$_{13}$ units. Initially we investigate IH(H$_2$)$_{13}$, a recently synthesised and exceptionally hydrogen-rich compound [Phys.\ Rev.\ B \textbf{97}, 024111 (2018)], which although is dynamically stable, is not recoverable. This structure consists of freely rotating 12 H$_2$ molecules in an icosahedral arrangement, enclosing another freely rotating H$_2$ molecule. Using these results, we search for alternative $X$(H$_2$)$_{13}$ materials and find that $X=\ $HCl, HBr, Xe are all dynamically stable. These compounds, and the icosahedral (H$_2$)$_{13}$ supermolecule, are all stable at similar volumes, suggesting appropriately--sized cavities could hold hydrogen supermolecules even at low pressure.
    
\end{abstract}
\maketitle

%%%%%%%%%%%%%%%%%%%%%%%%%%%%%%%%%
%%%%%%    Introduction      %%%%%
%%%%%%%%%%%%%%%%%%%%%%%%%%%%%%%%%
\section{introduction}
Hydrogen-rich compounds have attracted a lot of attention because of their potential applications for fuel cell storage\cite{jena2011materials} and for high-$T_c$ superconductors \cite{ashcroft2004hydrogen}. In particular, high pressure has been an effective thermodynamic variable in the synthesis of such compounds exemplified by recent experimental and theoretical breakthroughs. \cite{drozdov2015conventional,einaga2016crystal,ashcroft2004hydrogen,ashcroft1968metallic,jena2011materials,pickard2006high,martinez2009novel,errea2016quantum,zhang2015phase,zhong2012structural,liu2016crystal,eremets2008superconductivity,errea2015high,binns,peng2017hydrogen}.

The developments in high-pressure diamond anvil cells have pushed the
frontiers, with pressures up to 200~GPa now obtained routinely for most 
materials. However experimental studies involving hydrogen prove both problematic and costly, through the premature failure of the anvils due to hydrogen diffusion and embrittlement. Computational structure searches based on density functional theory
have now assumed an important role as a screening method in finding materials 
of potential interest as targets for synthesis\cite{pickard2011ab,wang2012calypso},
and first suggested very large hydrogen fraction compounds as a route to
high-$T_c$ superconductivity or energy storage\cite{zurek2009little} 

Recently, Binns \emph{et al.} synthesized\cite{binns} HIH($_2$)$_{13}$, a
compound with a number of remarkable features. 
At over 96\% stoichiometric ratio, it has the highest known
molecular hydrogen content of any compound. While this stoichiometry suggests
IH$_{27}$ to be a solid solution, X-ray diffraction
shows a primitive cubic lattice of iodine atoms\cite{binns}. The material 
can be regarded as a 1:1 stoichiometric compound comprising iodane
molecules and so-called (H$_2$)$_{13}$ supermolecules. Assuming that
each H$_2$ is spherical, as is the case for phase I of solid Hydrogen, three obvious possibilities present
themselves for the supermolecule: cuboctahedron, icosahedron or
dynamically spherical.  Icosahedral symmetry would be 
curious, because the fivefold symmetry is incompatible with
the cubic structure. The structure suggested by simulation has a
doubled unit cell along \{111\}, with two icosahedral supermolecules in opposite
orientations: this more complex structure is fully compatible with
existing experimental X-ray (which cannot detect the hydrogen positions), 
and Raman experimental data.

Hydrogen-storage materials face an intrinsic trade-off between
strongly binding hydrogen (atomic or molecular) for stability and the
energy required to release H$_2$ for applications. The (H$_2)_{13}$ units are
especially attractive because they store a large amount of hydrogen in
a weakly bound molecular form.  Other hydrogen-rich materials such as methane
hydrates or ammonium borohydride require energy to break strong bonds
between the hydrogen and other
elements\cite{somayazulu1996high,jena2011materials}.

It is interesting to consider whether the supermolecules may be
recovered to ambient conditions in any material.  HI(H$_2)_{13}$ decomposes
on depressurization, but this may be because the HI molecule itself is
unstable to decomposition to H$_2$ and I$_2$ at ambient
conditions\cite{binns2017synthesis}. Since the nature of the binding
in this material remains unknown, it is unclear whether this lack of
recoverability is a necessary consequence of the bonding type, or if
other similar materials might be recoverable.  Replacing the iodane
molecule with something stable is an obvious first step, e.g. other
hydrogen halides, large metallic atoms, noble gases or small polar
molecules like chloromethane.

% Density functional theory provides an accurate description of the
% electronic structure in most materials, and various structure search
% method have been very successful in discovering new
% phases.\cite{pickard2011ab,wang2012calypso} However, the physics of
% hydrogen-rich compounds is unusually challenging because of the
% importance of quantum effects on the nuclei.  Even in the classical
% nucleus approximation, standard structure search methods are unable to
% tackle materials containing rotating hydrogen molecules.  The reason
% for this is that an H$_2$ molecule in the J=0 rotational ground state
% is a spherical object, whereas in structure searches, it is a linear
% diatomic molecule.  The requirement for the molecule to point in a
% particular direction results in spurious symmetry-breaking 
% an issue which remains unresolved in structure searches.  
% Moreover,
% structure searches do not consider configurational entropy, which at
% extremely high hydrogen fraction will mean that solid solutions
% compete with crystalline energy minima\cite{peng2017hydrogen} at any
% finite temperature.
% 
% These central problems of symmetry-breaking and configurational entropy 
% can be addressed using {\it ab initio} 
% molecular dynamics simulations (AIMD).  These are more computationally
% expensive than the relaxations used in structure search, but they do
% also incorporate fully-anharmonic temperature effects.  Moreover,
% various projection-methods have been devised which enable
% spectroscopic quantities to be calculated, permitting direct
% comparison with experiment.

Here, we show that density functional theory (DFT) and \emph{ab initio} molecular
dynamics (AIMD) produce an accurate description of the known HI(H$_2)_{13}$ compound and analysis its behaviour under depressurization. Using these results, we investigate other potential compounds that could exhibit this hydrogen supermolecules and find HBr(H$_2)_{13}$, HCl(H$_2)_{13}$ and Xe(H$_2)_{13}$ to all be dynamically stable. The unprecedented stability of these compounds suggests that (H$_2$)$_{13}$ may be exhibited in a variety of systems and if recoverable, could have potential interest in hydrogen storage applications.

%%%%%%%%%%%%%%%%%%%%%%%%%%%%%%%%%
%%%%%%    Methods           %%%%%
%%%%%%%%%%%%%%%%%%%%%%%%%%%%%%%%%
\section{Methods}
DFT calculations were performed using the \textsc{castep} planewave
pseudopotential package\cite{clark2005first}, using norm-conserving
pseudopotentials, a minimum of $2\times 2\times 2$ k-points and an
energy cutoff of 1200~eV.  Temperature and anharmonic effects were
accounted for by performing Born-Oppenheimer AIMD, using a 0.5~fs
timestep. AIMD runs were typically 3~ps long.  This is necessarily far
too short of experimental timescales, but much longer than any
characteristic oscillation in the system, such that any soft-phonon or
martensitic transition could easily occur.  It is also long enough to
observe the free rotation of H$_2$ molecules.  We used a $2\times
2\times 2$ supercell with eight supermolecular units (224 atoms).

In all cases, the heavy molecules form a time-averaged structure close
to simple cubic, so for ease of comparison throughout, we always
report the lattice parameter associated with this lattice. This is
also the only structural variable that can be determined from X-ray
diffraction because the H$_2$ molecules are essentially invisible to X-rays.

The choice of exchange correlation functional is a highly debated
topic in DFT studies of hydrogen.  In a benchmarking studies on
hydrogen the BLYP functional\cite{blyp-LYP,blyp-B} was shown to give
results closest to Quantum Monte Carlo
calculations\cite{clay2014benchmarking,azadi2013fate}, for transition
pressures and vibrational frequencies, due mainly to the correct
asymptotic behaviour at high charge density gradient \cite{azadi2017role}. 

\begin{figure}
\includegraphics[width=0.5\textwidth]{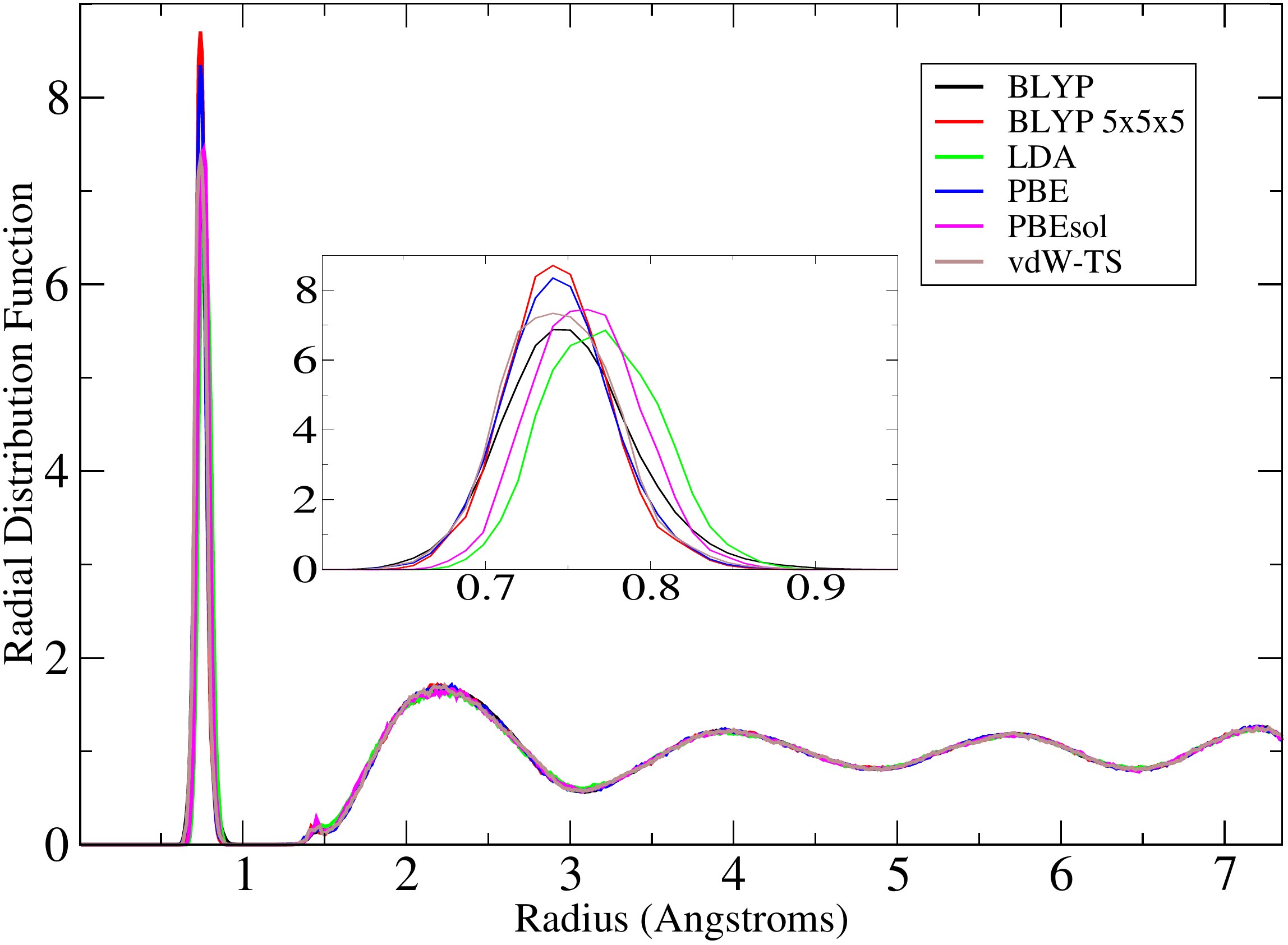}
\caption{All-atom RDF in HBr(H$_2)_{13}$ at
  lattice parameter 5.25\AA, calculated with a variety of
  exchange-correlation functionals. The small feature at 1.5~\AA\, is 
  the HBr bond.  
  The calculated pressures are 16.7~GPa (BLYP,\cite{blyp-LYP,blyp-B} 
  $5^3$ k-point grid), 20.5~GPa (BLYP,\cite{blyp-LYP,blyp-B}) 16.4~GPa 
  (PBE\cite{PBE}) 11.3~GPa (LDA\cite{LDA}), 13.3~GPa(PBEsol\cite{PBESOL}) 
  and 15.3~GPa(PBE+TS\cite{vdw-DF11}).\\  
  (\emph{inset}) Detail for the first peak, showing that LDA and PBEsol 
  give noticeably longer bondlengths commensurate with their lower 
  pressures.    
  MD simulations are NVT at 300~K, equilibrated for 1ps, then averaged 
  over 1.5ps and contain eight supermolecules.
\label{fig:rdf}}
\end{figure}

Van der Waals interactions are not well accounted for in traditional 
semilocal exchange-correlation functionals. These, however, could be 
important holding the hydrogen molecules together.  
 It has been shown that explicit
nonlocal treatment of the van der Waals interaction is of secondary
importance for hydrogen at pressures beyond 200~GPa\cite{azadi2017role}.  
% We mainly use BLYP here
The calculations shown here are performed with BLYP unless otherwise stated, 
with other exchange-correlation
functionals\cite{PBE,PBESOL,LDA,vdw-DF11} used for comparison in some
configurations (Figure~\ref{fig:rdf}).  As expected, we find %no
%significant difference between the Tkatchenko-Scheffler functional,
%which includes van der Waals explicitly, and the other forms which
%include it implicitly via the fit to the free electon gas or 
%molecular properties.  
Tkatchenko--Scheffler van der Waals corrections produce no significant
differences over the other semilocal functionals.
 The radial distribution function (RDF) shown in Fig.~\ref{fig:rdf}
shows that, at a given volume, the structural details are effectively 
independent of functional.
There is a bigger discrepancy in the computed DFT pressure. BLYP gives 
by far the largest values, and consequently a much higher ``pressure'' 
at given density compared with other functionals or with experiment.  
Pressure is a derived quantity in DFT, whereas volume is directly input, 
so it is reasonable to compare calculations at the same density.

%%%%%%%%%%%%%%%%%%%%%%%%   FIGURE 1   %%%%%%%%%%%%%%%%%%%%%%%%%%%%%
%%%%%%%%%%%%%%%%%%%%%%%%%%%%%%%%%%%%%%%%%%%%%%%%%%%%%%%%%%%%%%%%%%

In AIMD at 300~K, we observe that the hydrogen forms well defined
molecules with a high-frequency, Raman-active stretching mode
(vibron).  Molecules behave as mildly-inhibited rotors, so we expect their
Raman signature to be that of a quantum rotor (roton) at low pressure,
and a rotating harmonic oscillator (libron) at higher pressures.  This
also means that the molecules exhibit the correct spherical symmetry
and that, over time, the indistinguishability of the two nuclei is
respected. It is not clear to what extent the single molecule angular
momentum $J$ is a good quantum number, but in any case fully quantum
treatment with ortho-- and para--hydrogen is not currently possible.
Below 100K, molecular rotation stops, orientations are frozen into a
rotational glass.

Quantum nuclear effects are neglected in classical MD, and there are
several distinct places to consider where this can be
problematic\cite{ackland2015appraisal}.
\begin{itemize}
\item The high frequency H$_2$ molecular vibration is thermally
  excited in classical MD, but essentially not in the quantum case.  It adds a
  spurious contribution to the heat capacity which is therefore not
  reliable, and a zero-point energy contribution to the total
  energy. However, these contributions are essentially the same in all
  structures, and so cancel out in the free energy differences which
  determine stability.
\item The volume-dependence of the zero-point motion makes a
  contribution to the pressure.  This is significant at pressures
  above 200~GPa where the H$_2$ vibron weakens
  significantly\cite{ackland2015appraisal}.  At lower pressures, it is
  of order 1~GPa, close to the uncertainty due to exchange-correlation
  functional.
\item The zero-point motion means that there is already some energy in
  the bond, so breaking it would be easier if quantum nuclear effects
  were considered.  However, in our simulations we do not see
  significant bond breaking.
\item Non-interacting hydrogen molecules behave as quantum rotors with
  energy $\hbar^2J(J+1)/2I$, where $I$ is the moment of inertia.  The
  first excited state has energy around 170~K (H$_2$) or 85~K (other
  molecules), so the thermal excitation implied by our 300~K classical
  simulations are reasonable. Furthermore, the classical free rotor
  behaviour is a good representation of the spherically-symmetric
  $J=0$ ground state.  Thus we expect our supermolecule calculations
  to correctly describe the symmetry, just as classical MD of solid
  hydrogen gave an accurate qualitative prediction for the melting
  point of the ``quantum rotor'' phase\cite{bonev2004quantum}. By contrast, cooling
  simulations which show a symmetry-breaking phase transition are
  likely to overestimate the transition temperature.
\item The requirement for antisymmetic wavefunctions means that the
  nuclear spin states of ortho-- and para-- hydrogen couple to even and
  odd $J$ states respectively. This causes a slow equilibration of the
  roton energy.  However, it only applies when $J$
  %, the angular momentum of a single molecule - %%% Already defined
  is a good quantum number.  We monitor the
  angular momentum autocorrelation and find that typically the
  classical rotor cannot complete a full rotation without
  decorrelating.  Thus the rotors are coupled, $J$ is not a good quantum
  number, only the wavefunction of the entire system need be
  antisymmetric, and the nuclear spin state does not affect the
  dynamics.
\end{itemize}

So we expect any predictions of room temperature structures to be
robust with regard to quantum nuclear effects

The DFT calculations give reliable indications of stability, but to
obtain further insight one must look beyond the standard outputs to
consider the nature of the bonding. Pure high pressure hydrogen
changes under pressure from normal molecular phases, to phases in
which the electron is squeezed out of the bond, making it weaker and
longer\cite{pickard2006high,martinez2009novel,liu2012room,liu2012quasi,magdau2013identification,magdau2013high,howie2014phonon,magdau2017simple}.

We have developed some techniques for imaging and analysing rotating
molecules.  All of these depend on being able to identify H$_2$ molecules.
To do this, we first identify the nearest neighbor of each iodine.
Then, we find the nearest neighbour of each remaining hydrogen
atom. Provided each atom is its neighbor's neighbor, this uniquely 
defines molecules.  This procedure works at all but a few
timesteps at the highest pressures and temperatures considered here,
when H-H bond reconstruction %breaking and making 
sometimes occurs. %, in all materials. 
Such transforming configurations are excluded from the statistics.
 
Once molecules are identified, we can define a length $r_k$ and angular
momentum $\mathbf{l}_k$ for each molecule $k$.  For the vibrons, we exploit the fact that
Raman activity involves an in-phase symmetric stretch of
all the molecules.  We sum the total length of all molecules and build
the velocity autocorrelation function
\[ 
\textrm{VACF}(t) =  \left\langle \frac{d}{dt}\sum_k r_k(t)-\frac{d}{dt}\sum_k r_k(0) \right\rangle
\]
% where $r_k$ is the intramolecular distance and $k$ runs over all molecules.  
and the autocorrelation of the angular momentum:
\[ 
\textrm{LACF}(t) = \left\langle \frac{d}{dt}\sum_k \mathbf{l}_k(t)-\frac{d}{dt}\sum_k \mathbf{l}_k(0)  \right\rangle.
\]

The peaks in the Fourier transform of VACF correspond to the Raman-active 
vibron modes\cite{magdau2013identification}.  where $\mathbf{l}_k$ is the 
angular momentum of molecule $k$ with respect to its centre of mass.
Interpretation of the LACF is more subtle.  If we had
non-rotating molecules, its Fourier transform would give the
librational mode frequencies.  For free rotors the quantized
energies are unrelated to the period of rotation and are simply
$J(J+1)\hbar^2/mr_k^2$.  In the rotor case, the LACF tells us for how
long the molecule rotates unhindered.

%%%%%%%%%%%%%%%%%%%%%%%%%%%%%%%%%
%%%%%%    Validation        %%%%%
%%%%%%%%%%%%%%%%%%%%%%%%%%%%%%%%%
\subsection{The known material HI(H$_2)_{13}$}

The only currently known material containing the (H$_2)_{13}$
supermolecule is HI(H$_2)_{13}$.
The material was synthesized
by chemical reaction at pressure inside a diamond anvil cell \cite{binns}. The
simple-cubic arrangement of the iodine atoms was observed by X-ray
diffraction, and the iodane and hydrogen molecules were identified by
their characteristic Raman vibron frequencies. The composition was
determined by matching the density to the known I$_2$ and H$_{2}$
equations of state. 
The atomic arrangement was deduced from MD simulations, which support 
free IH and H$_2$ rotors at 300~K. Binns \emph{et al.} demonstrated
the \emph{average} icosahedral symmetry of the supermolecule by plotting
the mean \emph{atomic} positions: for a free-rotating HI molecule, the mean H
position is close to the centre of the Iodine while for H$_2$ the mean
position of both atoms is the same: the molecular centre.  
%This enables us to demonstrate the \emph{average} icosahedral symmetry 
%of the supermolecule. (Fig.\ref{fig:EOS}, inset)

\begin{figure}
\includegraphics[width=0.5\textwidth]{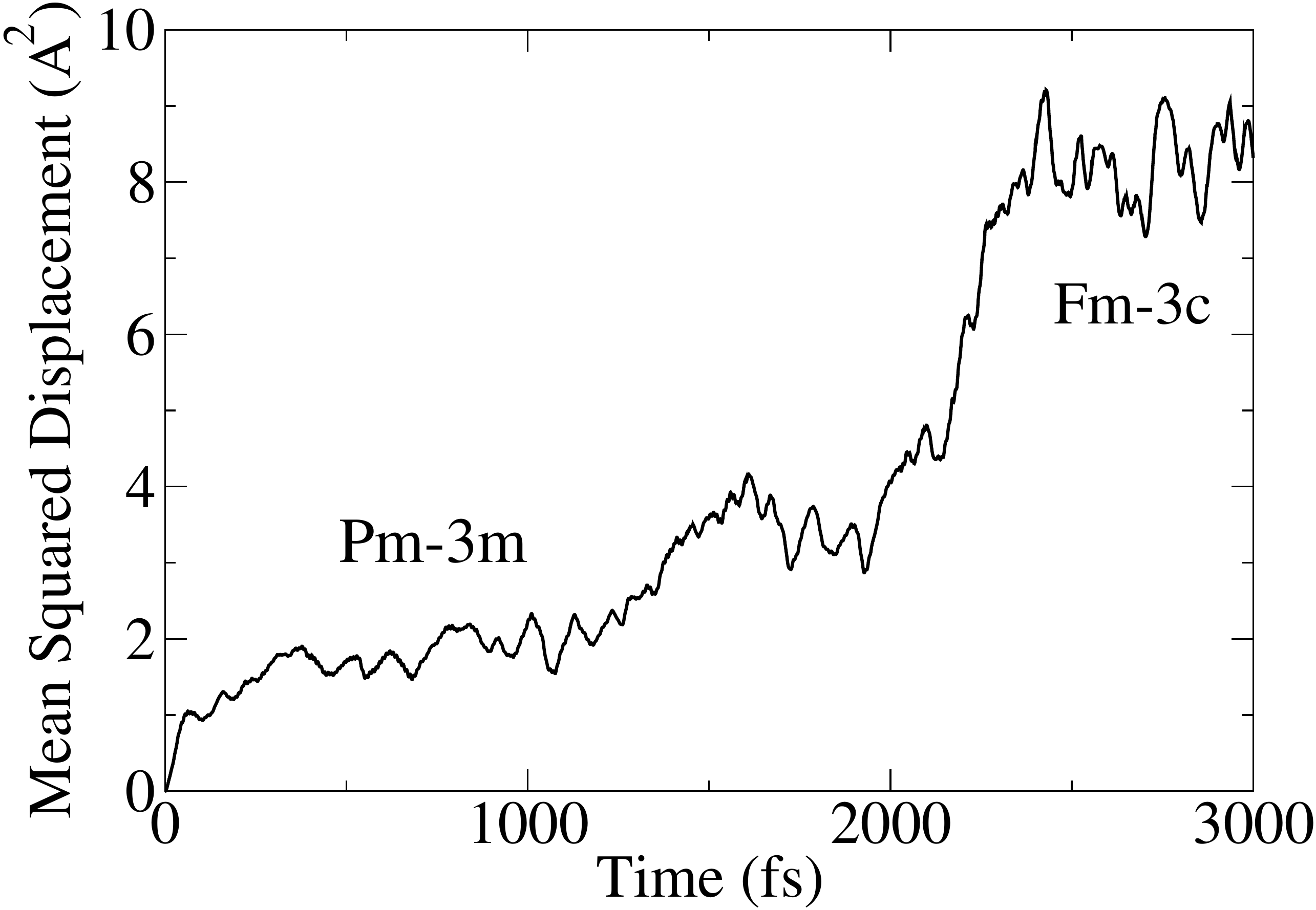}
\includegraphics[width=0.5\textwidth]{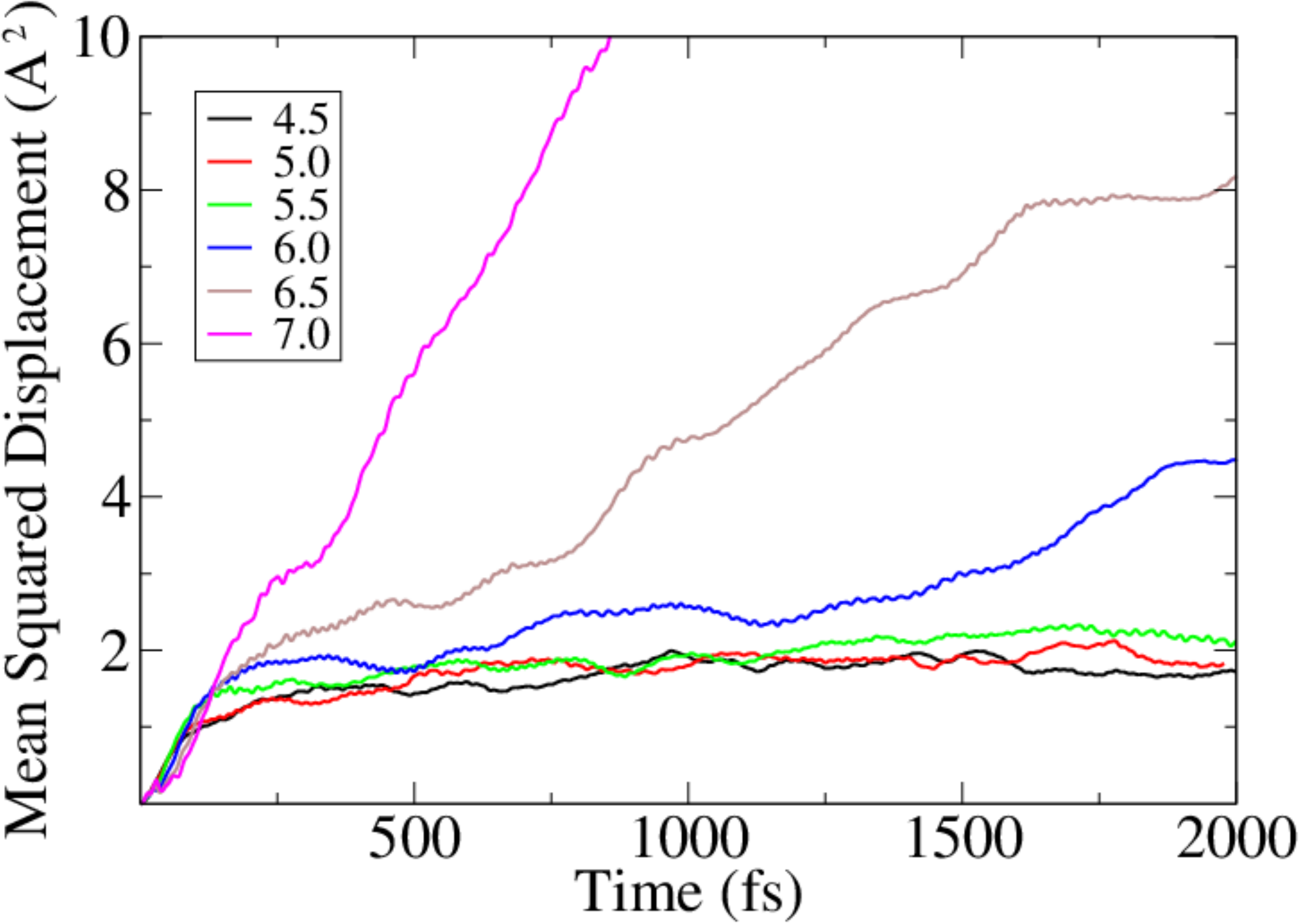}
\caption{\emph{(top)} MSD discontinuity at 2200~fs showing transformation of
  supermolecule in HI(H$_2)_{13}$ from cuboctahedral to icosahedral.
  \emph{(bottom)} MSD for HCl(H$_2)_{13}$ at 300~K and a
  range of lattice parameters, in \AA, as shown in the legend.
  The divergent MSDs correspond to low pressures below 7GPa.   %Calculations are done using BLYP.
\label{fig:msdHI}}
\end{figure}

In analogy with other binary AB$_{13}$ systems \cite{shoemaker1952interatomic,nordell1999linking,eldridge1993stability,schofield2005stability,jackson2007lattice} two clear candidate
structures involving cuboctahedral (P$m\overline{3}m$) and icosahedral (F$m\overline{3}c$) 
(H$_2)_{13}$ supermolecules emerge.
In order to find out the favoured structure, we carried out NPT ab-initio molecular 
dynamics simulation on these two phases at 30~GPa and 300~K, using the PBE functional.    
Using a primitive cubic cell, both P$m\overline{3}m$ (28 atoms) and
F$m\overline{3}c$ (56 atoms) remained stable over 3~ps.  Using a
larger cell with 224 atoms compatible with either structure,
simulations started with cuboctahedral supermolecules spontaneously
transformed to icosahedral ones.   
The molecular rearrangement is abrupt, and can be seen in the mean square 
displacement (MSD) shown in Fig.~\ref{fig:msdHI}.
This proves that the
F$m\overline{3}c$ structure, with two oppositely aligned icosahedral
supermolecules, is the stable structure.  
% Fig \ref{fig:msdHI}a shows the
% mean squared displacement (MSD) from this simulation, clearly showing the
% rearrangement, which goes from cuboctahedral to icosahedral.

%%%%%%%%%%%%%%%%%%%%%%%%   FIGURE 2   %%%%%%%%%%%%%%%%%%%%%%%%%%%%%
%%%%%%%%%%%%%%%%%%%%%%%%%%%%%%%%%%%%%%%%%%%%%%%%%%%%%%%%%%%%%%%%%

We then examined the possibility of recovering of the supermolecule.
This was done using a series of calculations at 300~K the NVT ensemble,
using the BLYP functional, and gradually reducing the density.
(Results are similar to the HCl(H$_2)_{13}$ data in Fig.~\ref{fig:msdHI}b, 
movies are available in SM).  
At lattice parameters greater than 7~\AA, the structure becomes unstable. 

We found that HI(H$_2)_{13}$ remains stable in a pressure range
between 6--100~GPa.  The equation of state is shown in Fig.~\ref{fig:EOS}.
We used the projection method to calculate the vibrational frequencies
of the Raman-active vibrons.  This showed that the two hydrogen
environments have very similar vibrational frequency, and we were
unable to distinguish a doublet.  The frequency is significantly
different from pure hydrogen at the same pressures.  The HI vibron
provides a well defined signature for iodane. We also did one run
setting the Iodine mass to that of hydrogen: although the dynamics are affected, in classical
Born-Oppenheimer MD, the free energy landscape is mass independent, so
this is equivalent to sampling for 200~ps --- still no structural
transformation occurred.

%%%%%%  Moved to Methods %%%%%%%
% We tested a number of other exchange-correlation functionals, and
% found that, the structures obtained are essentially independent of
% functional (Fig.\ref{fig:rdf}), only a small difference in
% bondlength. A much bigger discrepancy comes from the pressure
% calculated for a given structure. BLYP gives by far the largest
% values, and consequently a much higher ``pressure'' at given density
% compared with other functionals or with experiment.  Pressure is a
% derived quantity in DFT, whereas volume is directly input, so it is
% reasonable to compare calculations at the same density.

%%%%%%%%%%%%%%%%%%%%%%%%%%%%%%%%%
%%%%%%    New compounds     %%%%%
%%%%%%%%%%%%%%%%%%%%%%%%%%%%%%%%%
\subsection{Predicted compounds: \\ HBr(H$_2)_{13}$, HCl(H$_2)_{13}$ and Xe(H$_2)_{13}$}

Having established that molecular dynamics can determine the stability
of the supermolecule phases, we investigate a number of other possible
materials based on the same structure.  The dynamic nature of the
structure means that standard methods of ab initio structure searching
for static lattices cannot describe the symmetry.  So we resorted
again to molecular dynamics, to determine the metastability of the
structures.

At higher pressures, all three tested compounds were found to remain
stable throughout the simulations (up to 6~ps) at high pressures, but
at lower pressures the molecules began to exchange positions within
and between icosahedra, effectively forming a solution of HI, HBr or
HCl in fluid hydrogen.  We believe that the implication is that on
longer timescales the compound will be unstable at low density.  This
is shown by the divergent MSD of the hydrogen atoms in HCl(H$_2)_{13}$
for lattice parameters above 6.0~\AA\, (Fig.~\ref{fig:msdHI}b).  Similar
behavior is seen for both BLYP and PBE functionals at the same
density.  HCl(H$_2)_{13}$ is the most unstable of the materials here:
equivalent calculations for HI(H$_2)_{13}$, HBr(H$_2)_{13}$,
HI(H$_2)_{13}$, Xe(H$_2)_{13}$ respectively showed the breakdown in
stability occurs at increasing volume.  The Xe compound did not break
down, even at ambient pressure, and appears to be most stable.

\begin{figure}
\includegraphics[width=0.5\textwidth]{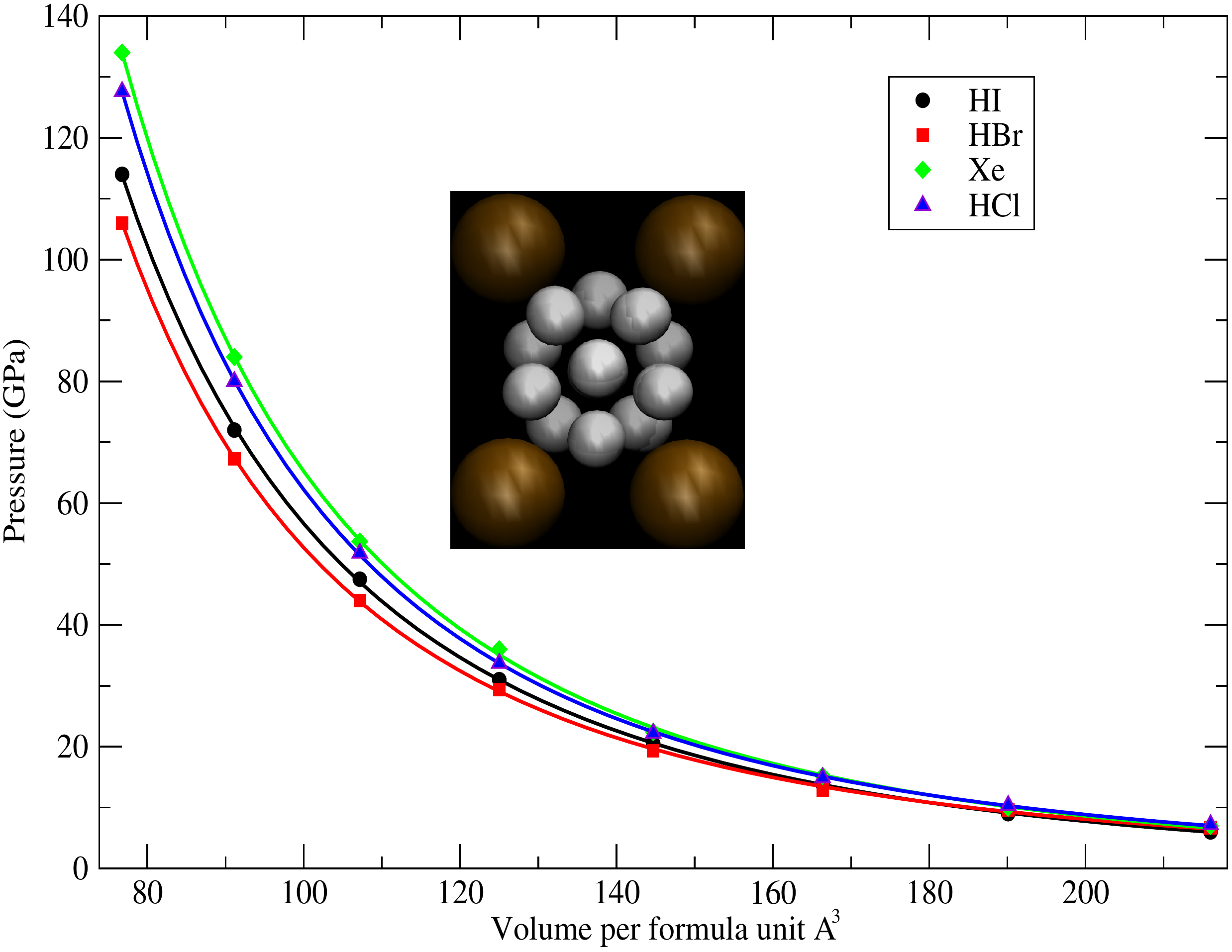}
\caption{Equations of state calculated using the BLYP functional.  The
  relative ordering of the curves is independent of the functional,
  but other functionals tested give lower values of pressure, so the
  systematic error in the calculated pressure is of order $\pm 5$GPa.
  Symbols are calculated pressure averages with statistical error less
  than symbol size. Lines show fits to Murnaghan equation of state.
  Inset show the mean atomic positions for IH(H$_2)_{13}$, emphasizing
  the fivefold axis of the supermolecule.
%I Fitted parameters = [  0.00000000e+00   5.19130780e+00   2.66695344e+00   3.01783459e+03]
%Br Fitted parameters = [  0.00000000e+00   5.49681001e+00   2.57148388e+00   2.90255385e+03]
%Xe Fitted parameters = [  0.00000000e+00   2.55167107e+00   2.68849276e+00   3.81222528e+03]
%Cl Fitted parameters = [  0.00000000e+00   1.41665076e+00   2.62553196e+00   4.59789457e+03]
\label{fig:EOS}}
\end{figure}

Comparative equations of state calculated with BLYP are shown in
Fig.~\ref{fig:EOS}. The three halide compounds show a clear trend in
the lattice parameter, with Xe falling between the HI and HBr
compounds. This scaling with molecular size is expected, much more
interesting is the comparison of the RDFs.  

%%%%%%%%%%%%%%%%%%%%%
%%%%%%  EOS   %%%%%%%
%%%%%%%%%%%%%%%%%%%%%

\begin{figure}
  \includegraphics[width=0.5\textwidth]{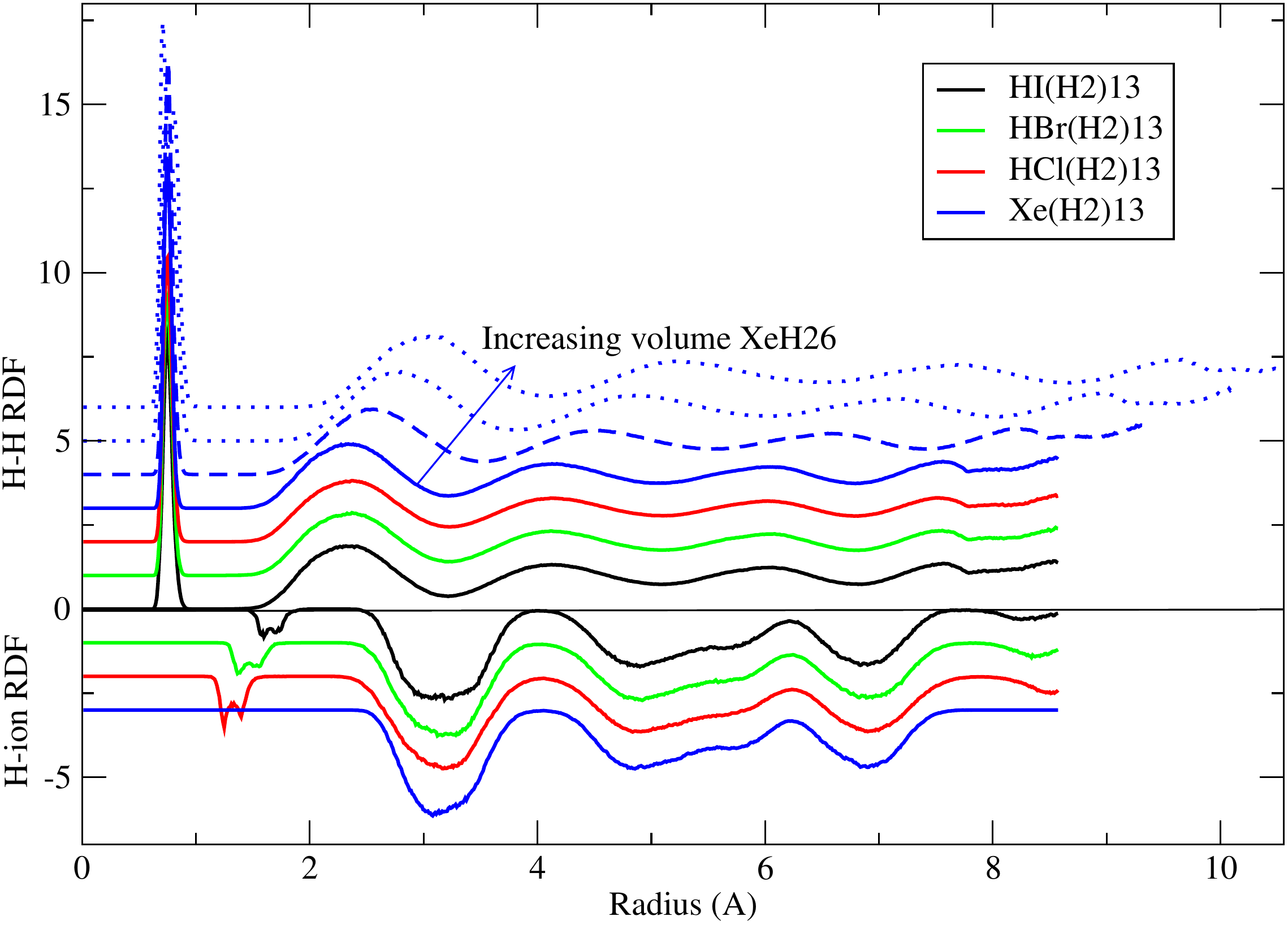}
  \caption{Radial distribution functions for hydrogen H-H (top)
  and H-X (bottom, X=Cl, I, Br, Xe) 
  for different compounds at the same lattice parameter (5.5\AA),
  and for the same compound, Xe(H$_2)_{13}$, at different densities (top lines in upper panel).
  \label{fig:rdf2}}
\end{figure}

Comparing different compounds at the same \emph{volume}
(Fig.~\ref{fig:rdf2}), the RDFs of the structures are essentially
identical, apart from the halide bondlength.  However, because of the
different equations of state, for different compounds at the same
\emph{pressure}, the supermolecule has very different intermolecular
distances.  The supermolecule behaviour is primarily determined by the
hydrogen density: the halide molecules provide an additional
\emph{``chemical pressure''} which determines the total pressure at a
given hydrogen density.

The equivalence of the hydrogen structure for all compounds implies
that the difference in pressure is mainly attributable to the
halide/Xe sublattice. 
Fig.~\ref{fig:rdf2} shows that the larger IH halide compound
is associated with the higher pressures.  However this ''molecular
volume'' argument is incompatible with the fact that the HI-H$_2$
intermolecular distances are the same as for HBr-H$_2$ and HCl-H$_2$.
Another possibility is that the materials with the higher permanent
dipole moment have stronger cohesion, due to dipole-dipole
correlations, and hence lower external pressure.  This explanation
works well for the halides, but is problematic with respect to the Xe
compound which has no dipole interactions.  Whatever the explanation,
it is clear that the pressure at which the supermolecule is stabilized
depends strongly on the other species.

%%%%%%%%%%%%%%%%%%%
%%%   RDF 2     %%%
%%%%%%%%%%%%%%%%%%%

%We carried out a calculation with only the (H$_2)_{13}$ molecules at a
%lattice parameter of 5.5\AA.  Unsurprisingly, the structure proved
%unstable, and the calculated pressure . This shows that compressing the hydrogen alone is a significant contribution to the total pressure.

%
%  Why is this here???
%

%%%%%%%%%%%%%%%%%%%%%%%%%%%%%%%%%
%%%%%%    Low temperature   %%%%%
%%%%%%%%%%%%%%%%%%%%%%%%%%%%%%%%%
\section{Low temperature phase transformations}

\begin{figure}
\includegraphics[width=0.5\textwidth]{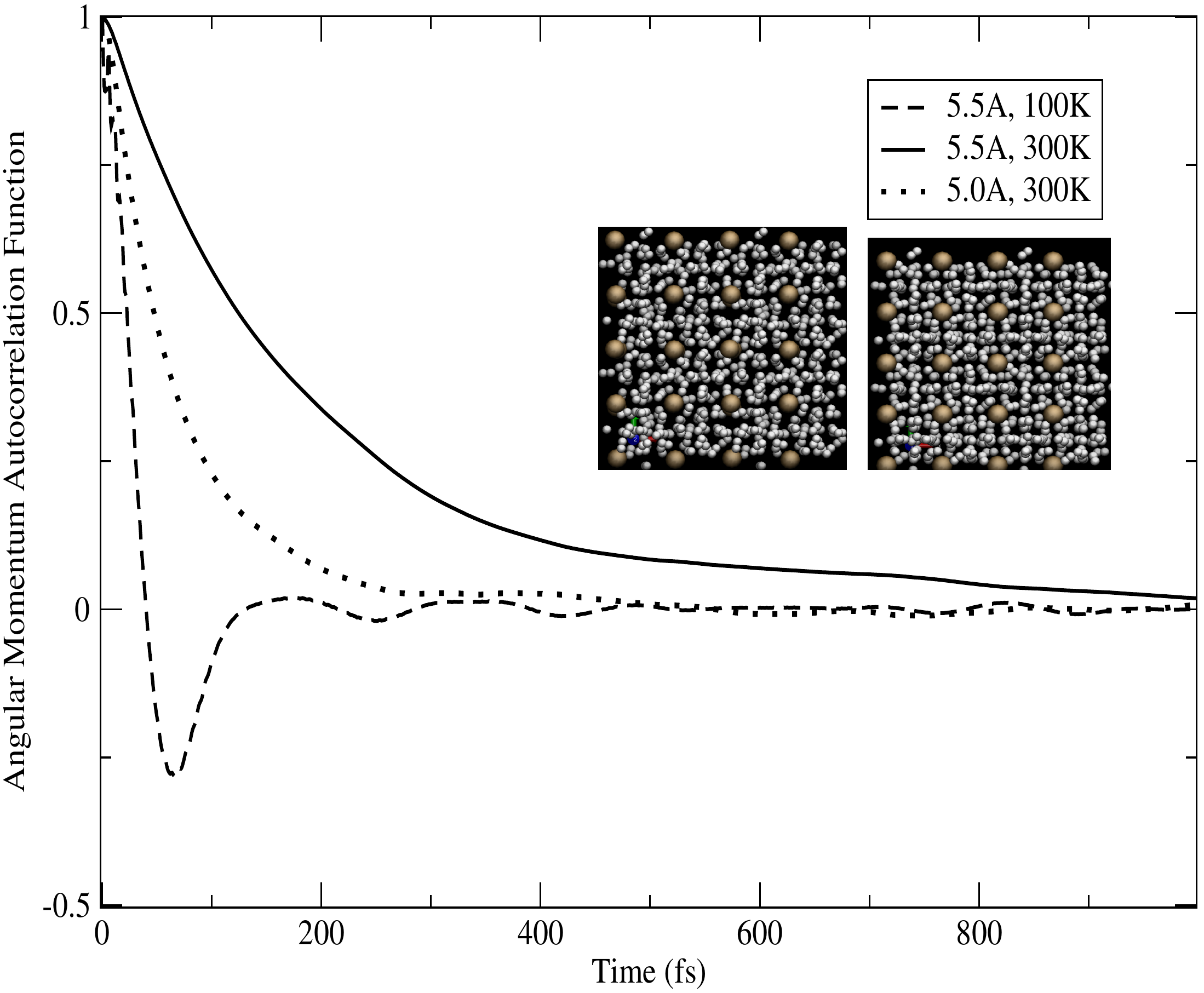}
\caption{%Angular momentum autocorrelation functions 
  LACF for Xe(H$_2)_{13}$ at 300~K (continuous) and 100~K (dashed, inset right), 
  with a 5.5 \AA\ lattice parameter. 
  The VACF at 300~K and $a=5.0$ \AA\ is shown for comparison.  
  Insets are MD snapshots at 5.5\AA,   showing the
  symmetry-breaking in the hydrogen (small spheres) sublattice on cooling.
\label{fig:AngCF}}
\end{figure}

Most free-rotor phases undergo a symmetry-breaking 
phase transformation when intermolecular forces are strong enough to stop the rotation.  In our simulations we observed that during
cooling to 100~K there is a transformation in the supermolecule to a
low temperature structure where the rotation is frozen and the molecules
librate rather than rotate.  This can be seen in animation (see SM)
and in the angular momentum correlation functions: Fig.~\ref{fig:AngCF}
show the results for Xe(H$_2)_{13}$ which is the most symmetric
structure on account of the monatomic Xe.

In the high-T symmetric phase the molecular orientation is disordered
and the rotation remains autocorrelated for a picosecond.  At lower-T,
the libration is shown by the negative region of the ACF, and
subsequent oscillation. The short correlation time indicates that the
single-molecule libron is not an eigenmode.  It is also notable that
the correlation time for the 300~K rotor is much reduced by pressure:
this will manifest experimentally as a broadening of the roton peak in
spectroscopy.  By 15~GPa (5.5~\AA) the autocorrelation function has a half-life
of less than 100~fs, which corresponds to less than a full rotation.

%%%%
%     Why is this here???
%%%%
The Xe compound best illustrates the broken symmetry of the
supermolecule (H$_2)_{13}$, since Xe is spherical.  The inset in
Fig.~\ref{fig:AngCF} shows a tendency for the molecules to align along
the cubic (100) axes.  The VACF-calculated Raman vibron spectrum shows two
discernible peaks, separated by over 100 wavenumbers, indicating that
there are two distinct H$_2$ environments, which would be clear
experimental signature of the symmetry-breaking on cooling.  In the halogen
compounds, the symmetry breaking is more complex because of the
dipole moment.

%%%%%%%%%%%%%%%%%%%
%%%     LACF    %%%
%%%%%%%%%%%%%%%%%%%

There has been some previous experimental and theoretical work on
xenon-hydrogen mixtures at lower hydrogen concentrations (90:10 gas
mixtures).  Somayazulu {\it et
  al}\cite{somayazulu2010pressure,somayazulu2015structure} found
compounds with stoichiometry Xe(H$_2)_7$ or Xe(H$_2)_8$.  From the
scattering pattern, they deduced the existence of covalent Xe-Xe dimer
bonds.  DFT structure search\cite{kaewmaraya2011theoretical} at these
compositions found candidate structures, but were unable to find any
Xe-Xe bonding and are compared directly with the experimental data.
As mentioned in the Methods section, ground-state structure search
inevitably produces symmetry-breaking from the orientation of the
hydrogen molecules.  Given the initial stoichiometries of these works,
it is unlikely that the Xe(H$_2)_{13}$ compound could have been found.

%%%%%%%%%%%%%%%%%%%%%%%%%%%%%%%%%
%%%%%%    Phonons           %%%%%
%%%%%%%%%%%%%%%%%%%%%%%%%%%%%%%%%
\section{Lattice Dynamics}

\begin{figure}[tbh]
\includegraphics[width=0.9\columnwidth]{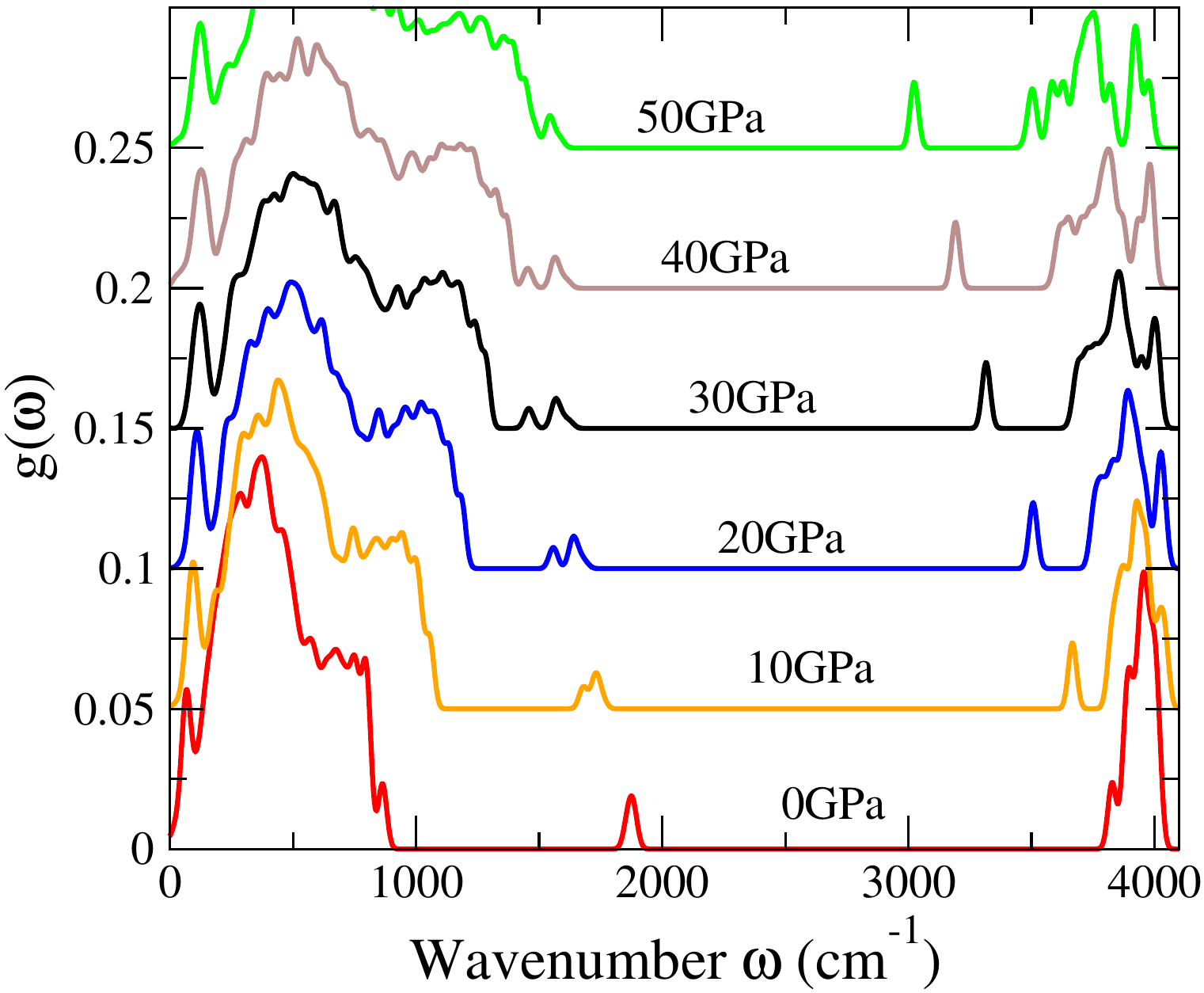}
\includegraphics[width=0.9\columnwidth]{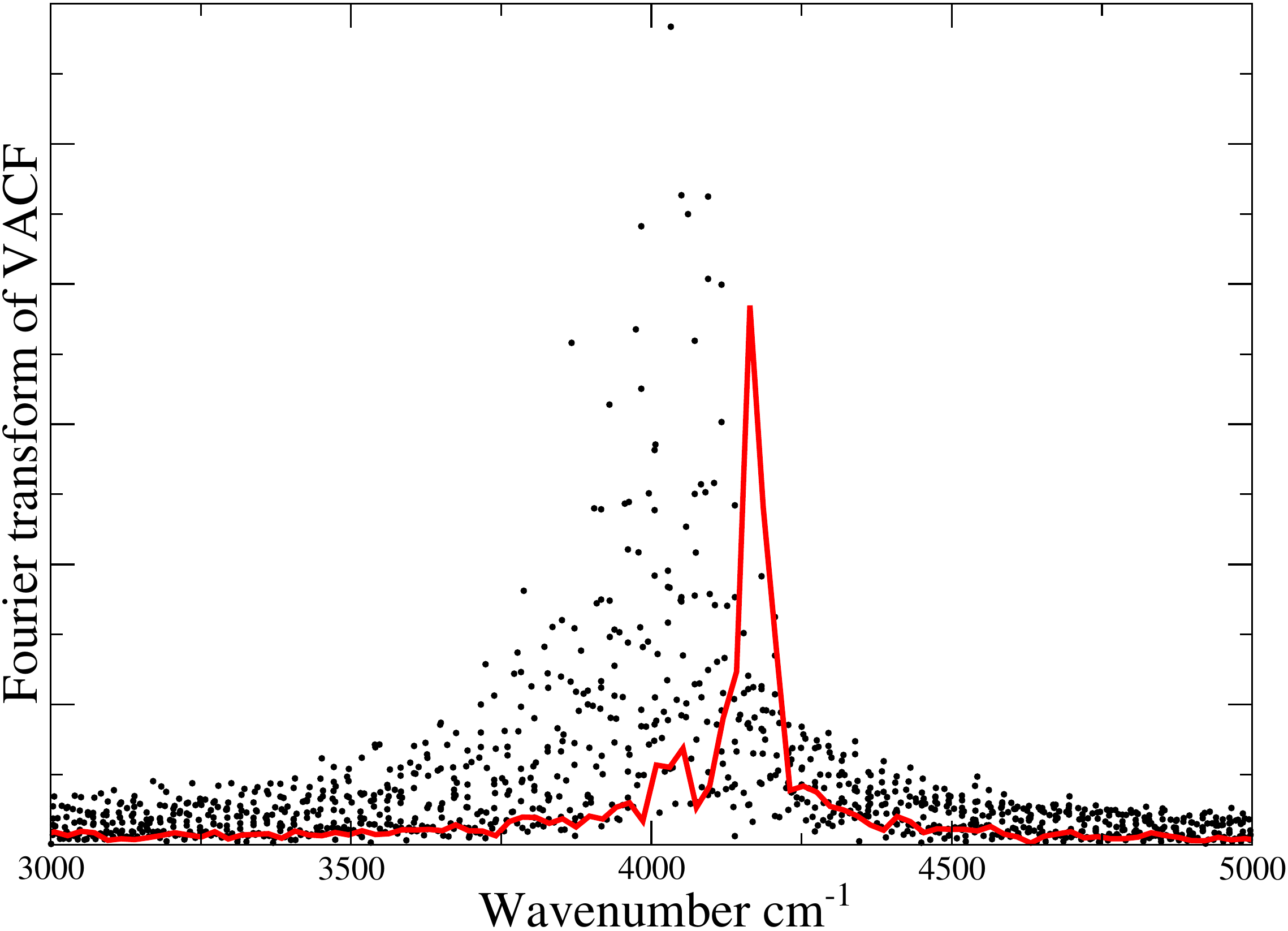}
\caption{(\emph{top}) Phonon DoS for a snapshot from HI(H$_2)_{13}$ MD
  relaxed from 300K at a range of pressures 0, 10, 20, 30, 40,
  50GPa. Calculation uses PBE 3x3x3 q-point grid and Gaussian
  broadening with FWHM of 40 cm$^{-1}$: since it is an MD snapshot no
  attempt is made to interpolate ``bands''. 
  (\emph{bottom}) Fourier transform
  of the VACF projected onto in-phase H$_2$ molecular
  vibrations\cite{magdau2013identification}, for HI(H$_2)_{13}$
  (black) and points from discrete Fourier transforms aggregated
  over all pressures, and a single calculation of pure hydrogen at
  20GPa (red).
\label{fig:phonon}}
\end{figure}

The Raman-active hydrogen vibron can be calculated from the MD using
the projection
method\cite{pinsook1999calculation,ackland2014efficacious}.
Fig~\ref{fig:phonon} shows the data for HI(H$_2)_{13}$.  Only a single
peak is evident, showing that the vibration of the central molecule in
(H$_2)_{13}$ has the same frequency as the others.  Moreover, this
frequency is significantly dependent on the functional used.  A
typical production run of 3~ps allows a discrete Fourier Transform to
sample every 10~cm$^{-1}$. This is sufficient to produce a peak around
4100~cm$^{-1}$, but not to discern any trend with pressure, and
so all pressures are shown together. Other functionals predict a somewhat
lower frequency: 3920~cm$^{-1}$ and 3860~cm$^{-1}$ for PBE and LDA respectively. 
 Thus despite the structural similarity, there is
strong dependence of vibron frequency on functional, correlated with a
variation of 2-3\% on H$_2$ bondlength.
Of these, the BLYP frequency is closest to the experimental value: 
4210$\pm$10~cm$^{-1}$. A similar calculation
for pure hydrogen at 20~GPa produces the red curve in Fig~\ref{fig:phonon},
and, despite the noise, it is clear that the hydrogen frequency is tens
of wavenumbers higher and probably sharper.  The hypothetical compounds mirror
this behaviour: the vibron frequency is always lower for the compound than 
for pure hydrogen.  We are not able to
discern any trend difference between the compounds, suggesting that
the supermolecular environment itself causes the shift.

%%%%%%%%%%%%%%%%%%%%%%%
%%%   Phonon DOS   %%%%
%%%%%%%%%%%%%%%%%%%%%%%

% For most materials, Raman intensity calculations using DFT lattice
% dynamics are straightforward, however, they are impossible in the
% limit of free-rotating molecules.  This is because lattice dynamics
% assumes the Raman signal comes from harmonic oscillators with atoms
% oscillating back and forth about an energy minimum at their mean
% position. The quantised energy is related to the frequency of this
% oscillation ($n+\frac{1}{2})\hbar\omega$.  For a 3-D rotating diatomic
% molecule, the atoms are located on a spherical shell centered on the
% molecular center of mass.  The average position of each atom is
% therefore at the molecular center, but in classical dynamics the atom
% is never found there.  Rotons are not spherical harmonic oscillators,
% and the quantised energy of a roton is not related to any classical
% frequency.  
%
% Furthermore, lattice dynamics assumes that a Raman signal correspond
% to a normal mode expressed in cartesian coordinates. This assumption
% breaks down with rotors. Consider for example a molecule pointing
% along the $x$ direction, the normal mode corresponding to the
% stretching mode involves atomic displacements along $\pm x$.  Suppose
% that later the molecule rotates to point along the $y$ direction; now
% the atomic displacements along $\pm x$ would correspond to molecular
% rotation, not stretching.  Thus Raman modes cannot be associated with
% cartesian normal-mode eigenvectors.

For classical crystals, Raman intensity calculations using DFT lattice
dynamics and perturbation theory are straightforward. However, the
small--amplitude approximation breaks down in the limit of
freely-rotating molecules: consider for example a molecule pointing
along the $x$ direction, the normal mode corresponding to the
stretching mode involves atomic displacements along $\pm x$.  Suppose
that later the molecule rotates to point along the $y$ direction; now
the atomic displacements along $\pm x$ would correspond to molecular
rotation, not stretching.  Thus Raman modes cannot be associated with
cartesian normal-mode eigenvectors.  Rotons are not spherical harmonic
oscillators, and the quantised energy of a roton is not related to any
classical frequency.

Although coupling between rotons and other Raman modes means one
cannot simply apply standard lattice dynamics, it is instructive to
make some attempt.  So we extracted a snapshot from the MD
calculation, relaxed it to the nearest zero-temperature enthalpy
minimum for a range of pressures, and calculated the phonon density of states 
(DoS) between 0 and 50~GPa. The results are shown in Fig.~\ref{fig:phonon}.  
Although this is not a
definitive representation of the structure, a comparison at different
pressures using the same snapshot is meaningful and  
% Lattice dynamics
% with PBE systematically underestimates the vibron
% frequency\cite{martinez2009novel,clay2014benchmarking,magdau2017infrared,ackland2015appraisal},
% however
the Phonon DoS shows a number of interesting features.

Firstly, the main effect of pressure is to broaden the vibron band,
and shift it to lower frequencies.  Secondly, the mode has much lower
frequency than the equivalent vibron in pure hydrogen (e.g. using identical 
settings the Raman mode for pure hydrogen\footnote{Using the
 P$ca2_1$ structure, a candidate for the ordered hydrogen Phase~II.},
at 30~GPa is 4292~cm$^{-1}$).  This is consistent with the
observation of Binns et al. who saw two vibrons and attributed the
higher frequency mode to pure hydrogen.\cite{binns} The H$_2$
molecule bondlength is around 0.75~\AA, compared to 0.736~\AA in pure hydrogen. 
It suggests that the H-H bond in the supermolecule is weaker, more like that in 
hydrogen Phase~III which occur at much greater pressure.

Our phonon calculation also shows a pronounced weakening of the IH
bond with pressure, from 2000 to 1500~cm$^{-1}$: again consistent with
experiment.  The reason for this can also be seen in by a detailed
analysis of the modes.  One mode in Fig.~\ref{fig:phonon},
corresponding to a single H$_2$ molecule, becomes detached from the
main vibron band and significantly weakens with pressure.
Interestingly, the molecule in question is \emph{not} the one in the
centre of the supermolecule, rather it is the one closest to the H ion
in the IH molecule.  By 50~GPa, its bondlength has stretched to
0.843~\AA, and the Mulliken bond population is  $\sim 25$\% lower than
other molecules.  Even so, this vibron has a much higher frequency then
the rotation of the IH molecule.  So it is reasonable to treat the IH
dipole orientation as fixed throughout many vibrational periods of the
vibron: in short, we believe that this bond weakening is real.  The
weaked H$_2$ bond would be difficult to detect by spectroscopy, because with a lifetime
equivalent to the rotation rate of the IH, it will have a linewidth of
hundreds of wavenumbers, not the arbitrarily-chosen 40~cm$^{-1}$ of the
figure.  However, as they rotate, all the IH will \emph{always} have a nearby
hydrogen, so the weakening of its bond is permanent and detectable.

Comparing the two phonon methods, we see that the projection method
gives a smaller range of vibron frequencies than the lattice dynamics,
as expected since they are only the Raman active modes. Using the full
VACF without projection gives a broader band and the low frequency
modes, similar to the lattice dynamics.  VACF also picks out the IH
vibron, but not the weakening H$_2$ mode: this strongly suggests that
the latter mode is an artifact of the snapshot used in the lattice dynamics.

%%%%%%%%%%%%%%%%%%%%%%%%%%%%%%%%%
%%%%%%    Discussion        %%%%%
%%%%%%%%%%%%%%%%%%%%%%%%%%%%%%%%%
\section{Discussion}

We have shown that the icosahedral (H$_2)_{13}$ supermolecule is a
unit from which to build hydrogen-rich crystal structures.  The
structure is similar to some AB$_{13}$ compounds, such as NaZn$_{13}$
and beryllium-rich intermetallics.  But whereas those are held
together by metallic bonding, (H$_2)_{13}$ is unique in having only
van der Waals or quadrupole interactions to stabilize it. An alternative reason for stability of AB$_{13}$
is observed in binary hard sphere mixtures and opals (packed SiO$_2$
spheres) because it has low entropy thanks to its dense packing.\cite{eldridge1993stability,jackson2007lattice}

The icosahedral, 13-member supermolecule has been calculated to be the
most stable molecular cluster dues to van der Waals interaction in
Lennard-Jonesium and in
hydrogen\cite{wales1997global,cuervo2006path,martinez2007theoretical}.
The motif cannot be extended to fill homogeneous 3D space, but the
current study demonstrates that by packing alongside other molecules,
extended structures can be built from this stable cluster.

In DFT calculations, the fundamental quantity is the volume.  Pressure
is a derived quantity, and the work emphasises the dependence of the
pressure on exchange-correlation functional.  In particular, the BLYP
functional, known to give the closest agreement with Quantum Monte
Carlo\cite{azadi2013fate,clay2014benchmarking} for dense hydrogen,
gives calculated pressures higher than other functionals or
experiments.  The robust conclusions of this paper are therefore
framed in terms of density rather than pressure.

The key question for understanding the prospects for recoverability is
whether the supermolecule is primarily stabilized by:
\begin{itemize} 
\item free energy of the hydrogen molecules themselves, 
\item interactions involving HI, Xe, HBr or HCl molecules,
\item the external pressure and efficient packing ($PV$).
\end{itemize} 
In the present cases, the evidence is in favor of packing efficiency:
although the HBr, HCl and HI molecule have dipole moments, the
structure persists when they are freely rotating, and the Xe compound
is unlikely to be stabilised by long - range Xe-Xe bonds. Furthermore,
the hydrogen structure is the same in all compounds at the same
lattice parameter, while the intermolecular distances within the
supermolecule are highly dependent on density. Explicit treatment of
van der Waals interactions is critical at low densities in hydrogen, but of
secondary importance at pressure\cite{azadi2017role}.  Here van der Waals interactions modelled by the
Tkatchenko-Scheffler correction to PBE produced the same
structures, further evidence that packing via the $PV$ term in the
enthalpy is more important.  There is no characteristic supermolecule
size.  Nevertheless, the pressure is dependent on the minority
species.

 The hydrogen molecules in the supermolecule show a number of
 behaviours typical of pure hydrogen at higher pressures.  In all
 cases the H$_2$ bonds are weaker than in pure hydrogen at equivalent
 pressures; this manifests both as longer bondlengths and lower
 vibrational frequencies.  Also, the supermolecule transforms to a
 broken-symmetry structure at low temperature and rotation becomes
 inhibited at higher pressure: features observed in pure hydrogen at
 higher pressures. Hence it is conceivable that on further pressure
 increase this supermolecular network might become electrically
 conducting.

If the molecular-spaced  H$_{26}$ cannot be recovered,
another possible route to hydrogen-storage utilizing  (H$_2)_{13}$ is 
a caged structure with cavities tuned to the size of the supermolecule.

Whether recoverable to ambient conditions or not, these supermolecules
compounds have novel and exceptional properties: The combination of
extreme stoichiometry and incorporation of fivefold symmetry within a
crystal structure are properties previously exhibited by fullerene
compounds.  The icosahedral supermolecule is stable in a range of
environments and will be the basis of a new class of ordered
hydrogen-rich compounds.

\begin{acknowledgments}
We would like to thank EPSRC for computing resources through the UKCP project, the ERC ``Hecate'' fellowship for GJA, and the Royal Society Wolfson Award.
\end{acknowledgments}

\bibliographystyle{apsrev}

\end{document}